\documentclass{IEEEtran}
\usepackage[USenglish,english]{babel}
\usepackage{amsmath}
\usepackage{algorithm}
\usepackage{algpseudocode}
\usepackage{booktabs}
\usepackage{csquotes}
\usepackage{caption}
\usepackage{xcolor}
\usepackage{subcaption}
\usepackage{tikz}
\usepackage{todonotes}
\usepackage{url}
\usepackage{hyperref}
\usepackage{xspace}
\usepackage{numprint}
\usepackage{flushend}
\usetikzlibrary{fit,matrix,positioning}

\renewcommand{\vec}[1]{\mathbf{#1}}
\newcommand{\mat}[1]{\mathbf{#1}}

\newcommand{\eg}{e.g.,\xspace}
\newcommand{\ie}{i.e.,\xspace}

\newif\iffullpaper
\fullpapertrue

\newcommand{\egc}{e.\,g.,\xspace}
\newcommand{\etal}{et al.\xspace}

\newcommand{\change}[1]{\textcolor{red}{#1}}
\renewcommand{\change}[1]{#1}

\npdecimalsign{.}
\npthousandsep{\,}

\newcommand{\constructioncombblasminspeedup}{$\numprint{1.68}$\space}
\newcommand{\constructioncombblasmaxspeedup}{$\numprint{2.59}$\space}

\newcommand{\constructiondcsrspeedup}{$\numprint{1.15}$\space}
\newcommand{\insertioncombblasminspeedup}{$\numprint{3.63}$\space}
\newcommand{\insertioncombblasmaxspeedup}{$\numprint{227.68}$\space}
\newcommand{\insertionctfminspeedup}{$\numprint{55.15}$\space}

\newcommand{\insertionpetscminspeedup}{$\numprint{460.83}$\space}

\newcommand{\updatecombblasminspeedup}{$\numprint{3.75}$\space}
\newcommand{\updatecombblasmaxspeedup}{$\numprint{263.57}$\space}
\newcommand{\updatectfminspeedup}{$\numprint{59.8}$\space}

\newcommand{\updatepetscminspeedup}{$\numprint{477.11}$\space}

\newcommand{\deletioncombblasminspeedup}{$\numprint{4.86}$\space}
\newcommand{\deletioncombblasmaxspeedup}{$\numprint{393.85}$\space}
\newcommand{\deletionctfminspeedup}{$\numprint{101.43}$\space}

\newcommand{\dynamicspgemmcombblasminspeedup}{$\numprint{3.41}$\space}
\newcommand{\dynamicspgemmcombblasmaxspeedup}{$\numprint{6.18}$\space}
\newcommand{\dynamicspgemmctfminspeedup}{$\numprint{11.73}$\space}

\newcommand{\dynamicspgemmpetscminspeedup}{$\numprint{5.2}$\space}

\newcommand{\generalspgemmcombblasminspeedup}{$\numprint{2.39}$\space}
\newcommand{\generalspgemmcombblasmaxspeedup}{$\numprint{4.57}$\space}
\newcommand{\generalspgemmctfminspeedup}{$\numprint{14.58}$\space}

\newcommand{\generalspgemmpetscminspeedup}{$\numprint{6.9}$\space}

\newcommand{\constructionsynthstrongspeedup}{$\numprint{10.85}$\space}

\usepackage{ifthen}
\newboolean{blinded}

\begin{document}

\title{Fast Dynamic Updates and Dynamic SpGEMM \\ on MPI-Distributed Graphs}

\author{
	\IEEEauthorblockN{
		Alexander van der Grinten\IEEEauthorrefmark{1},
		Geert Custers\IEEEauthorrefmark{2},
		Duy Le Thanh\IEEEauthorrefmark{1},
		and Henning Meyerhenke\IEEEauthorrefmark{1}
	}\\
	\IEEEauthorblockA{
		\IEEEauthorrefmark{1}Humboldt-Universität zu Berlin
		Berlin, Germany
	}\\
	\IEEEauthorblockA{
		\IEEEauthorrefmark{2}Delft University of Technology,
		Delft, Netherlands
	}
}

\maketitle

\begin{abstract}
Sparse matrix multiplication (SpGEMM) is a fundamental kernel used in many diverse application areas, both numerical and discrete. For example, many algebraic graph algorithms rely on SpGEMM in the tropical semiring to compute shortest paths in graphs. Recently, SpGEMM has received growing attention regarding implementations for specific (parallel) architectures. Yet, this concerns only the static problem, where both input matrices do not change. In many applications, however, matrices (or their corresponding graphs) change over time. Although recomputing from scratch is very expensive, we are not aware of any dynamic SpGEMM algorithms in the literature.

In this paper, we thus propose a batch-dynamic algorithm for MPI-based parallel computing. Building on top of a distributed graph/matrix data structure that allows for fast updates, our dynamic SpGEMM reduces the communication volume significantly. It does so by exploiting that updates change far fewer matrix entries than there are non-zeros in the input operands. Our experiments with popular benchmark graphs show that our approach pays off. For batches of insertions or removals of matrix entries, our dynamic SpGEMM is substantially faster than the static algorithms in the state-of-the-art competitors CombBLAS, CTF and PETSc.

\end{abstract}

\section{Introduction}
%
Sparse general matrix-matrix multiplication, usually denoted by SpGEMM, is a fundamental computational
kernel for many applications in various areas -- for instance scientific computing (such as algebraic multigrid~\cite{STUBEN2001331}), machine learning with sparse DNNs~\cite{DBLP:conf/hpec/KepnerAG0MRS19}, and
data analytics (such as clustering~\cite{doi:10.1137/040608635}).
Besides its use in numerical contexts, numerous graph algorithms (such as triangle counting~\cite{yacsar2018fast})
make use of SpGEMM due to the correspondence between graphs and matrices~\cite{DBLP:books/siam/11/KG2011,DBLP:conf/hpec/KepnerABBFGHKLM16}.
Two other popular applications are shortest paths with multiple sources and graph contraction~\cite{DBLP:journals/corr/abs-2002-11273}.
Thus, our results apply to both the matrix and the graph perspective as well as for numerical and discrete problems.

In fact, we focus on algorithms and data structures for \emph{dynamic} SpGEMM in this paper.
Note that dynamic problems are very common in data analytics applications -- 
for example, consider the continuously changing inputs in areas such as recommender systems,
online social networks, or time-dependent mobility networks.
Moreover, the use of dynamic SpGEMM is also conceivable for deep learning on changing/growing data sets.
Frequently recomputing analytics results (or SpGEMM as one step in the workflow) after
changes in the input data is typically very time-consuming or even infeasible for large inputs.
It is thus desirable to use a dynamic algorithm that can update the previous result in a more
cost-efficient manner. Also, numerous large-scale applications require (or at least benefit from) 
the speed and memory of HPC platforms. 
Yet, despite growing research interest in optimizing static 
SpGEMM for various architectures and as part of MPI-based tools (see Section~\ref{sec:rel-work} for a brief
overview), no dynamic approach for SpGEMM has been published so far -- neither sequential nor parallel/distributed.

\paragraph*{Outline and Contributions}
We thus first propose an MPI-parallel data structure for fast dynamic updates to sparse matrices and graphs
(Section~\ref{sec:data-structure}). This data structure stores the adjacency matrix in a 2D distribution over the processes. 
Based on this data structure, we propose a parallel dynamic SpGEMM algorithm in Section~\ref{sec:dynamic-distributed-spgemm}.
More precisely, we provide two variants, one for algebraic updates (update addition equals the one in the SpGEMM semiring) and one for general updates.
The main idea behind the algorithm is to limit the communication between processes at the potential
cost of a mild increase in other operations. To achieve this goal, we change the way block submatrices
are communicated (and aggregated) between processes while using our dynamic matrices.
Our experimental results (see Section~\ref{sec:experiments}) show that our approach pays off.
A faster redistribution yields much better insertion, update, and deletion times for the data structure
compared to the state-of-the-art tools CombBLAS, CTF and PetSC.
Regarding SpGEMM, the dynamic algorithm for algebraic updates (which performs better than the general
one) is between \dynamicspgemmcombblasminspeedup and
\dynamicspgemmcombblasmaxspeedup times faster on average than a static SpGEMM computation with CombBLAS (the best competitor) --
depending on the batch size with up to some \emph{millions} of non-zero updates.
Finally note that related work is presented in Section~\ref{sec:rel-work}, preliminaries and notation in 
Section~\ref{sec:prelim}, while concluding remarks are made in Section~\ref{sec:conclusion}.

\section{Related Work}
\label{sec:rel-work}
%
Sparse graph and matrix computations in general have received considerable research attention in the last decades.
One more recent appealing idea is to express graph computations by linear algebraic operations~\cite{DBLP:books/siam/11/KG2011},
which has led to the GraphBLAS initiative and standard~\cite{DBLP:conf/hpec/KepnerABBFGHKLM16,DBLP:conf/hpec/MattsonYMBM17}.
Compared to sparse matrix-vector multiplication (SpMV), the more challenging SpGEMM has been covered to a lesser extent in the literature so far -- as pointed out by Winter \etal~\cite{DBLP:conf/ppopp/WinterMZSS19}.
Recently, however, the optimization of static SpGEMM algorithms
for specific parallel architectures is on the rise, \egc for multithreaded CPUs~\cite{DBLP:conf/icppw/NagasakaMAB18,DBLP:conf/spaa/GuMEA20} 
GPUs~\cite{DBLP:conf/ppopp/WinterMZSS19}, CPU/GPU combinations~\cite{DBLP:conf/hpec/EllisR19,DBLP:journals/tpds/XieTLS22}, and other accelerators~\cite{DBLP:conf/icppw/NagasakaMAB18}.
A possible reason for this spike could be the use of SpGEMM in deep learning with sparse DNNs, as described in Ref.~\cite{DBLP:conf/hpec/KepnerAG0MRS19}.
Rather than providing a comprehensive overview over static SpGEMM, our description focuses on efforts 
for MPI -- as only these are directly comparable to our approach. For more details, the interested reader is referred to a recent
systematic literature review on SpGEMM~\cite{DBLP:journals/corr/abs-2002-11273}.

MPI-based frameworks for sparse graph computations and/or sparse linear algebra include CombBLAS~2.0~\cite{9470983},
Trilinos~\cite{trilinos-website}, CTF~\cite{SOLOMONIK20143176}, LAMA~\cite{DBLP:books/sp/17/BrandesSS17}, and PETSc~\cite{petsc-efficient}. 
In our experiments, we compare against recent versions of CombBLAS, CTF and PETSc.

As pointed out by Bulu\c{c} \etal~\cite{DBLP:books/siam/11/BulucGS11},
the most common data structures for sparse matrices and graphs in the context of algebraic operations
are variants of compressed sparse row (CSR) or column (CSC).
This also holds for the parallel case, where the distribution of the matrix over the processes is an important aspect.
The two most common distributions are $1$D (vertex-based in graphs terms) and $2$D (edge-based)~\cite{DBLP:books/siam/11/BulucG11},
but also $2.5$D and $3$D are in use~\cite{doi:10.1137/15M104253X}.
In particular, for highly irregular inputs, a 2D distribution fares better than 1D due to better load 
balancing~\cite{DBLP:conf/dimacs/BulucM12}. A $3$D implementation with multiple levels of parallelism~\cite{doi:10.1137/15M104253X}
was shown to outperform previous SpGEMM codes for MPI parallelism.
Parallel distributions can lead to hypersparse (sub-)matrices on some processes, 
for which doubly compressed versions of CSR/CSC are known~\cite{DBLP:books/siam/11/BulucG11}. 
In the same work, Bulu\c{c} and Gilbert~\cite{DBLP:books/siam/11/BulucG11} also propose sparse SUMMA as their
``algorithm of choice'' for static parallel SpGEMM due to its flexibility in terms of matrix size.
SUMMA is also used as part of the $3$D algorithm by Azad \etal~\cite{doi:10.1137/15M104253X} (see above).
We argue in Section~\ref{sec:dynamic-distributed-spgemm} why sparse SUMMA is not well suited for a dynamic approach.

The potential of (batch-)dynamic algorithms -- algebraic or not -- for other graph/matrix problems 
has been shown in numerous recent works, among many others for core maintenance with matchings~\cite{8357916}, 
all-pair shortest paths and betweenness centrality~\cite{DBLP:conf/wea/BergaminiMOS17}, 
as well as various other centrality measures~\cite{DBLP:conf/esa/GrintenBGBM18,DBLP:conf/ipps/Riedy16}.


\section{Preliminaries and Notation}
\label{sec:prelim}

Our work is built on top of MPI. We denote the number of MPI processes
by $p$. Within each MPI process, we also make use of shared memory
(\ie OpenMP) parallelism.

We consider matrices over arbitrary semirings;
for an introduction of matrix computations over
semirings (in the context of graphs),
we refer the reader to Ref.~\cite{DBLP:conf/hpec/KepnerABBFGHKLM16}.
Common examples include the $(+, \cdot)$ semiring,
the $(\land, \lor)$ semiring over Boolean values,
or the $(\min, +)$ semiring (which is often used
in shortest path algorithms).
We usually use the symbol $0$ to refer to the neutral element of the semiring.

Matrices are denoted by $\mat{A}$, $\mat{B}$, $\mat{C}$ and similar.
The $(u, v)$-th entry of $\mat{A}$ is denoted by $a_{u,v}$.
Our framework will distribute blocks of matrices to each
MPI process in a 2D process grid of size $\sqrt p \times \sqrt p$.
In this context, $\mat{A}_{i,j}$ refers to the block of $\mat{A}$
that is located on MPI process $(i, j)$, where $i, j \in \{1, \ldots, \sqrt p\}$.
We assume that matrices (\eg input matrices for SpGEMM)
are \emph{sparse}, \ie that $nnz(\mat{A}) \ll n \cdot m$
for an $n \times m$ matrix $\mat{A}$.
Some matrices that arise in our algorithms will be
\emph{hypersparse}, \ie $nnz(\mat{A}) \ll n$ and $nnz(\mat{A}) \ll m$.
Like most other sparse matrix frameworks, we differentiate between
\emph{structural} (non-)zeros and \emph{numerical} (non-)zeros.
In particular, an entry $a_{u,v}$ of a matrix $\mat{A}$ is considered
to be structurally non-zero if $a_{u,v}$ is present in our sparse matrix
data structure, even if $a_{u,v} = 0$.
In particular, a structural non-zero can still have the numerical
value of zero.
The term \emph{non-zero} will always be used for structural
non-zeros throughout this paper.
Structural zeros (\ie entries of the matrix that are not present in
our data structures), are always implicitly equal to
the (additive) neutral element of the semiring (\eg $\infty$ for $(\min, +)$).

\emph{SpGEMM} (sparse generalized matrix multiplication)
is the problem of computing $\mat{C} = \alpha \mat{A} \mat{B} + \beta \mat{X}$,
where $\mat{A}$ and $\mat{B}$ may optionally be transposed.
Since the main challenge of any SpGEMM algorithm is computing the matrix
multiplication $\mat{A} \mat{B}$, most of this paper will be focused
on that computation, although we also give details about handling
transposition in Section~\ref{sec:algo-transpose}.


\section{Data Structures for Dynamic Distributed Graphs}
\label{sec:data-structure}
Our framework stores each matrix in a fully distributed way.
We employ a 2D distribution of the matrix, i.e., each MPI process stores
a block of the matrix.
Like other distributed sparse matrix/graph frameworks (such as CombBLAS~\cite{9470983}),
we support square process grids of size $\sqrt p \times \sqrt p$
for this purpose.

To store these blocks locally within each process, we distinguish between
\emph{dynamic} and \emph{static} matrices.
Dynamic matrices support efficient in-place operations
(such as insertions, deletions, matrix addition or other element-wise transformations).
To store dynamic matrices, we use the DHB data structure~\cite{dhb}
which is based on adjacency arrays, together
with a per-row hash table that maps column indices
to locations in the adjacency array.
This data structure allows us to quickly discover whether
a given index pair $(i, j)$ has a non-zero value;
it also enables us to efficiently update
a local entry of the matrix in $\mathcal{O}(1)$
expected time.

We store static matrices in a CSR data structure.
For various hypersparse static matrices that arise
in our algorithms, we use a
\emph{doubly compressed} CSR (= DCSR)
matrix instead, similarly to the DCSC matrix that is
used by CombBLAS~\cite{DBLP:conf/ipps/BulucG08}.
DCSR (or DCSC)
further decrease the memory overhead of a sparse matrix
by storing only row pointer for non-zero rows (or columns, respectively).
Doubly compressed storage layouts come
at the cost of not being able to directly index (\ie in $\mathcal{O}(1)$ time)
into a specific row (or column) anymore.
However, they can substantially decrease communication
volume when hypersparse matrices need to be communicated.
Since none of our algorithms needs to index into a static
CSR or DCSR (\ie no search for an index is ever necessary),
we do not sort these storage layouts in any way
(and we also do not maintain hash tables for these layouts).

\subsection{Dynamic Updates}

As a \emph{dynamic update}, we consider the modification
of a non-zero entry of an existing matrix.
An update can consist of the insertion of a new non-zero,
the deletion of an existing non-zero,
or the change of the numerical value of a matrix entry.
In many cases, such updates can be represented
by adding an \emph{update matrix} to the original matrix;
however, this is not true in general semirings such as $(\min, +)$,
where the $\min$ operator can only decrease (but not increase)
the value of non-zeros of the matrix.
We assume that MPI processes
can generate updates independently and without knowledge of
the distribution of data across the MPI process grid.

While our algorithms and data structures for dynamic updates are
different compared to the state of the art in MPI-distributed sparse
matrix processing, not many changes at the interface
are necessary to support dynamic updates.
To insert new entries into a matrix $\mat{A}$, we first
build an update matrix $\mat{A^*}$ that contains
exactly the new values of all non-zeros that should
be updated in $\mat{A}$.
Afterwards, we either add $\mat{A^*}$ to $\mat{A}$
(if the update can be represented by addition),
or we call a $\Call{merge}{\mat{A}, \mat{A^*}}$ procedure that replaces
all values $(i, j)$ of $\mat{A}$ by their corresponding
entry in $\mat{A^*}$ if the $(i, j)$-th entry of $\mat{A^*}$
is non-zero.
To delete entries, we support a
$\Call{mask}{\mat{A}, \mat{A^*}}$ procedure that
removes all entries $(i, j)$ from $\mat{A}$
for which $\mat{A^*}$ is non-zero.
Similar operations (optimized for the static case)
are already present in most state-of-the-art sparse matrix frameworks.

The construction of $\mat{A^*}$ involves communication
to redistribute data across MPI processes.
After $\mat{A^*}$ is constructed, however, all dynamic update
operations (\ie matrix addition, \textsc{merge}, \textsc{mask})
operate only on local blocks of the matrices;
no communication is involved.
All update operations can be implemented efficiently if $\mat{A}$
is stored as a dynamic matrix and $\mat{A^*}$ is stored in
DCSR layout.
To ensure that these matrices are indeed stored in these layouts,
our framework requires the user to mark dynamic matrices
and update matrices appropriately.

\subsection{Distribution of Update Matrices across MPI Processes}
\label{app:distr-updates}

Since MPI processes can generate updates independently
and without knowledge of the data distribution,
we need to be able to quickly
redistribute a set of updates to the MPI process that
stores the matrix block that is affected by the update.
Our redistribution uses straightforward techniques;
however, we found our algorithm to outperform
existing MPI-parallel implementations in practice
(in particular, this holds for implementations that rely
on comparison-based sorting and a single \textsc{AllToAll},
see Section~\ref{sec:experiments}).
To redistribute updates, we represent them as tuples $(i, j, x)$,
where $(i, j)$ is the index pair that should be updated
to a new value $x$. Our redistribution routine takes
an array of such tuples as input on each MPI process.
We first exchange tuples across the rows of the process
grid. In a second step, we exchange these tuples
(which are now on the correct row of the process grid)
across the columns of the grid. An \textsc{AllToAll}
communication call is used to transfer the data.
Before each \textsc{AllToAll}, we use counting sort
to group the tuples by their destination rank.
While \textsc{AllToAll} is expensive in general,
our two-step approach ensures that this call only affects
$\sqrt p$ processes. Likewise, our counting sort
has to consider only $\sqrt p$ buckets.
We use OpenMP parallelism to efficiently insert
updates into local dynamic matrices
(\ie into adjacency arrays and associated hash tables).
To enable this procedure, we use counting sort
to group updates $(i, j)$ according to $(i \mod T)$, where $T$ is the number
of (shared-memory) threads.
Afterwards, we handle updates with different $(i \mod T)$ in parallel.

\paragraph*{Analysis}
Let $\mathrm{nnz}$ be the number of input tuples that need
to be redistributed. We assume that all processes
initially have $\mathrm{nnz}/p$ of these tuples.
Applying a random permutation to the input tuples
(before invoking redistribution)
allows us to assume that load is evenly distributed
even after redistribution~\cite{DBLP:journals/corr/SolomonikH15}.
Due to the \textsc{AllToAll} call,
our redistribution has a communication latency of $\sqrt p$
and bandwidth requirements of $\mathcal{O}(\mathrm{nnz}/p)$,
since each process sends at most as much data as it locally has.
We can assume that inserting into a local dynamic matrix
(\ie into an adjacency list + hash table)
can be done in expected amortized $\mathcal{O}(1)$ time.
Hence, the (expected amortized) local computation time
is $\mathcal{O}(\mathrm{nnz}/p + \sqrt p)$,
\ie it is dominated by our counting sort that groups non-zeros
by their destination process.


\section{Dynamic Distributed SpGEMM}
\label{sec:dynamic-distributed-spgemm}
We now consider the problem of dynamically
updating the result of an SpGEMM operation.
In particular, we want to compute $\mat{C} = \mat{A} \mat{B}$, where $\mat{A}$
is a $n \times k$ matrix and $\mat{B}$ is a $k \times m$ matrix.
Since our work concerns the dynamic variant
of this matrix multiplication,
we assume that $\mat{A}$ and $\mat{B}$
have been updated by inserting and/or removing non-zero entries.
These updates result in new matrices $\mat{A'}$ and $\mat{B'}$, respectively.
The \emph{dynamic matrix multiplication} problem now consists
of computing $\mat{C'} = \mat{A'} \mat{B'}$, given
the result $\mat{C}$ of the multiplication before updates were applied.
We consider two cases of the dynamic matrix multiplication problem:

\begin{itemize}
	\item \textbf{Algebraic updates.} We say that $\mat{A'}$ is obtained
		by \emph{algebraic} updates if $\mat{A'} = \mat{A} + \mat{A^*}$
		is the sum of $\mat{A}$ and an update matrix $\mat{A^*}$,
		where the sum uses the same addition operator as the semiring
		that is used in the SpGEMM.
		If $\mat{A'}$ and $\mat{B'}$ are obtained by algebraic updates,
		we can exploit the distributive property of matrix multiplication
		to compute $\mat{C'}$ as
		$\mat{C'} = (\mat{A} + \mat{A^*}) (\mat{B} + \mat{B^*})
		= \mat{A} \mat{B} + \mat{A} \mat{B^*} + \mat{A^*} \mat{B} + \mat{A^*} \mat{B^*}
		= \mat{C} + \mat{A^*} \mat{B'} + \mat{A} \mat{B^*}$.
		Throughout the remainder of this paper, we refer to the
		last two terms of the latter expression as
		$\mat{C^*}$, \ie
		\begin{equation}
			\mat{C^*} := \mat{A^*} \mat{B'} + \mat{A} \mat{B^*}.
			\label{eq:cupdate}
		\end{equation}
		The main goal of our algorithm for algebraic updates is
		to compute $\mat{C^*}$ without performing extensive communication
		(\ie without communicating $\mat{B'}$
		or $\mat{A}$).\footnote{In the formulation of this paper, we always
			compute $\mat{C^*}$ as $\mat{A^*} \mat{B'} + \mat{A} \mat{B^*}$.
			Our algorithms can also be rewritten to perform the computation
			as $\mat{C^*} = \mat{A^*} \mat{B} + \mat{A'} \mat{B^*}$.
			Both expressions are equally suitable for our algorithms;
			they also yield identical algorithmic complexities.}

		We note that if the semiring that is used for the matrix multiplication
		is a ring, this covers all possible updates since
		$\mat{A^*}$ can simply be computed as $\mat{A^*} = \mat{A'} - \mat{A}$
		in rings (but not in general semirings).
	\item \textbf{General updates.} If the algebraic update condition does not hold,
		we say that $\mat{A'}$ and $\mat{B'}$ are obtained by \emph{general}
		updates.
		General updates can update matrices in ways that are
		\enquote{incompatible} with the semiring used for the matrix multiplication.
		For example, they can set values to $0$ in the $(\lor, \land)$ semiring
		or increase values of non-zeros of the matrix in the $(\min, +)$ semiring.

		Even in the general case, we can define update matrices
		$\mat{A^*}$ and $\mat{B^*}$ that contain only the non-zeros
		of $\mat{A'}$ and $\mat{B'}$ that have changed compared to $\mat{A}$ and $\mat{B}$.
		For our algorithms, only the structure (and not the numerical values)
		of $\mat{A^*}$ and $\mat{B^*}$ will be relevant.
		Given these matrices, we can compute $\mat{C^*}$ as in Eq.~(\ref{eq:cupdate}).
		While $\mat{C'} \neq \mat{C} + \mat{C^*}$ in the general case,
		$\mat{C'}$ can still only differ from $\mat{C}$ at entries
		that are non-zero in $\mat{C^*}$.

		Note that deletions of non-zeros in the input matrices can be handled
		in our algorithm for general updates by removing the corresponding
		entries from $\mat{A'}$ or $\mat{B'}$ while adding a structural non-zero
		to $\mat{A^*}$ or $\mat{B^*}$
		to indicate that the corresponding entries in the input matrices have
		changed.
\end{itemize}

In both the algebraic and the general case, we assume that update matrices
(\ie $\mat{A^*}$ and $\mat{B^*}$) are hypersparse
(while $\mat{A'}$ and $\mat{B'}$ are in general sparse, but not hypersparse).
Our algorithm works for any number of non-zeros in $\mat{A^*}$ and $\mat{B^*}$;
however, our analysis of compute- and communication time relies on hypersparsity.


\subsection{Algorithm for Algebraic Updates}
\label{sec:spgemm-algebraic}

\tikzset{process/.style={minimum width=4mm,minimum height=4mm},
	process matrix/.style={left delimiter={[},right delimiter={]},row sep=2mm,column sep=1mm}
}

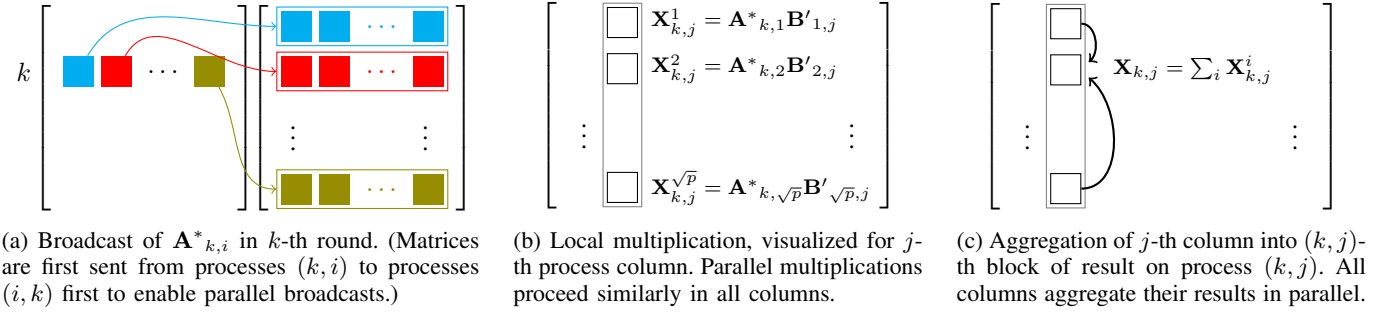
\begin{figure*}[t]%
\begin{subfigure}[t]{.35\textwidth}
\centering
\begin{tikzpicture}
\matrix[process matrix] (A) {
	\node[process] {};
		& \node[process] {};
		& \node[process] {};
		& \node[process] {}; \\
	\node[fill,cyan,process] (Ak1) {};
		& \node[fill,red,process] (Ak2) {};
		& \node[process] {$\ldots$};
		& \node[fill,olive,process] (Ak4) {}; \\
	\node {\phantom{$\vdots$}}; & & & \node {\phantom{$\vdots$}}; \\
	\node[process] {};
		& \node[process] {};
		& \node[process] {};
		& \node[process] {}; \\
};
\matrix[process matrix,right=5mm of A] (Abcast) {
	\node[fill,cyan,process] (Abcast11) {};
		& \node[fill,cyan,process] {};
		& \node[process,cyan] {$\ldots$};
		& \node[fill,cyan,process] (Abcast14) {}; \\
	\node[fill,red,process] (Abcast21) {};
		& \node[red,fill,process] {};
		& \node[red,process] {$\ldots$};
		& \node[red,fill,process] (Abcast24) {}; \\
	\node {$\vdots$}; & & & \node {$\vdots$}; \\
	\node[fill,olive,process] (Abcast41) {};
		& \node[fill,olive,process] {};
		& \node[olive,process] {$\ldots$};
		& \node[fill,olive,process] (Abcast44) {}; \\
};

\node[left=8mm of Ak2] {$k$};

\node[draw,cyan,fit=(Abcast11)(Abcast14),inner sep=.5mm] (Abcast1) {};
\node[draw,red,fit=(Abcast21)(Abcast24),inner sep=.5mm] (Abcast2) {};
\node[draw,olive,fit=(Abcast41)(Abcast44),inner sep=.5mm] (Abcast4) {};

\draw[->,cyan] (Ak1) to [out=60,in=180] (Abcast1);
\draw[->,red] (Ak2) to [out=60,in=180] (Abcast2);
\draw[->,olive] (Ak4) to [out=300,in=180] (Abcast4);
\end{tikzpicture}
\caption{Broadcast of $\mat{A^*}_{k,i}$ in $k$-th round.
	(Matrices are
	first sent from processes $(k, i)$ to processes
	$(i, k)$ first to enable parallel broadcasts.)}
\label{subfig:algo-broadcast}
\end{subfigure}\hfill%
\begin{subfigure}[t]{.30\textwidth}
\centering
\begin{tikzpicture}
\matrix[process matrix] (Aloc) {
	\node[process] {};
		& \node[draw,process] (Aloc12) {};
		& \node[process] {\hspace{18mm}\phantom{$\ldots$}};
		& \node[process] {}; \\
	\node[process] {};
		& \node[draw,process] (Aloc22) {};
		& \node[process] {};
		& \node[process] {}; \\
	\node {$\vdots$}; & & & \node {$\vdots$}; \\
	\node[process] {};
		& \node[draw,process] (Aloc42) {};
		& \node[process] {};
		& \node[process] {}; \\
};

\node[draw,gray,fit=(Aloc12)(Aloc42),inner sep=.5mm] {};

\node[right=.5mm of Aloc12,align=center] {\footnotesize
	$\mat{X}^1_{k,j} = \mat{A^*}_{k,1} \mat{B'}_{1,j}$};
\node[right=.5mm of Aloc22,align=center] {\footnotesize
	$\mat{X}^2_{k,j} = \mat{A^*}_{k,2} \mat{B'}_{2,j}$};
\node[right=.5mm of Aloc42,align=center] {\footnotesize
	$\mat{X}^{\sqrt p}_{k,j} = \mat{A^*}_{k,\sqrt p} \mat{B'}_{\sqrt p,j}$};
\end{tikzpicture}
\caption{Local multiplication, visualized for $j$-th process column.
	Parallel multiplications proceed similarly in all columns.}
\label{subfig:algo-local}
\end{subfigure}\hfill%
\begin{subfigure}[t]{.30\textwidth}
\centering
\begin{tikzpicture}
\matrix[process matrix] (Agath) {
	\node[process] {};
		& \node[draw,process] (Agath12) {};
		& \node[process] {\hspace{18mm}\phantom{$\ldots$}};
		& \node[process] {}; \\
	\node[process] {};
		& \node[draw,process] (Agath22) {};
		& \node[process] {};
		& \node[process] {}; \\
	\node {$\vdots$}; & & & \node {$\vdots$}; \\
	\node[process] {};
		& \node[draw,process] (Agath42) {};
		& \node[process] {};
		& \node[process] {}; \\
};

\node[draw,gray,fit=(Agath12)(Agath42),inner sep=.5mm] {};

\node[right=3mm of Agath22] {\footnotesize $\mat{X}_{k,j} = \sum_i \mat{X}^i_{k,j}$};

\node[right=-1.5mm of Agath22] (Agather) { };
\draw[->,thick] (Agath12) to [out=0,in=40] (Agather);
\draw[->,thick] (Agath42) to [out=0,in=320] (Agather);
\end{tikzpicture}
\caption{Aggregation of $j$-th column into $(k,j)$-th
	block of result on process $(k,j)$.
	All columns aggregate their results in parallel.}
\label{subfig:algo-aggregate}
\end{subfigure}%
\caption{Communication pattern of MPI-parallel dynamic SpGEMM algorithm for algebraic updates.
Between broadcast and aggregation, all MPI processes $(i, j)$ compute $\mat{A^*}_{k,i} \mat{B'}_{i,j}$ locally.
Figures depict computation of $\mat{A^*} \mat{B'}$; computation of $\mat{A} \mat{B^*}$
proceeds similarly.}
\end{figure*}

\begin{algorithm}[t]
\caption{MPI-parallel dynamic SpGEMM, algebraic updates}
\label{algo:dist-mm}
\begin{algorithmic}
\State \Comment{Executes in parallel on all MPI processes $(i,j)$}
\State send $\mat{A^*}_{i,j}$ to process $(j, i)$, receive $\mat{A^*}_{j,i}$
\State send $\mat{B^*}_{i,j}$ to process $(j, i)$, receive $\mat{B^*}_{j,i}$
\For{$k = 1, \ldots, \sqrt p$}
	\State broadcast $\mat{A^*}_{k,i}$ over $i$-th process row
	\State broadcast $\mat{B^*}_{j,k}$ over $j$-th process column
	\State $\mat{X}^i_{k,j} \gets \mat{A^*}_{k,i} \mat{B'}_{i,j}$ \Comment{local}
	\State $\mat{Y}^j_{i,k} \gets \mat{A}_{i,j} \mat{B^*}_{j,k}$ \Comment{local}
	\State aggregate $\mat{X}^i_{k,j}$ to $\mat{X}_{k,j}$ on process $(k, j)$
	\State aggregate $\mat{Y}^j_{i,k}$ to $\mat{X}_{i,k}$ on process $(i, k)$
\EndFor
\State $\mat{C'}_{i, j} \gets \mat{C}_{i,j} + \mat{X}_{i,j} + \mat{Y}_{i,j}$ \Comment{local}
\end{algorithmic}
\end{algorithm}

As detailed above, we can compute $\mat{C'} = \mat{C} + \mat{C^*}$
in the algebraic case, with $\mat{C^*}$ as in Eq.~(\ref{eq:cupdate}).
The standard algorithm to evaluate the matrix multiplications
$\mat{X} := \mat{A^*} \mat{B'}$ and $\mat{Y} := \mat{A} \mat{B^*}$
in MPI-distributed frameworks is the SUMMA algorithm
(= \emph{scalable universal matrix multiplication}).
SUMMA performs $\sqrt p$ rounds. In each round, it
broadcasts blocks of the left-hand side of the matrix multiplication
over the rows of the process
grid, and blocks of the right-hand side over the columns of the grid.
This procedure ensures that process $(i, j)$ will receive
exactly the blocks $(i,k)$ of the left-hand side and $(k,j)$ of the right-hand side
that are required to form block $(i, j)$ of the result
\ie exactly the block of the output matrix that is supposed to reside on process $(i, j)$.
Due to the way data is broadcasted in SUMMA, the aggregation of partial results
into block $(i, j)$ of the result is entirely local.
SUMMA has an efficient communication pattern if both left-hand and right-hand sides
have similar numbers of non-zeros. However, in our case (e.g., when computing
$\mat{A^*} \mat{B'}$), one of the matrices (namely, $\mat{A^*}$)
is expected to have far fewer non-zeros than the other matrix,
since $\mat{A^*}$ represents updates that usually affect only a small subset of $\mat{A}$.

Algorithm~\ref{algo:dist-mm} depicts the pseudocode of our
algebraic algorithm for dynamic SpGEMM.
Instead of relying on the usual SUMMA algorithm, our
algorithm can be seen as a combination of two passes of input-stationary
SUMMA~\cite{DBLP:journals/siamsc/SchatzGP16}, without materializing intermediate results.
In particular, our algorithm avoids broadcasting blocks of $\mat{A}$ and $\mat{B'}$.
This comes at the cost of an additional non-local aggregation step.
However, since we expect $\mat{A^*}$, $\mat{B^*}$,
and $\mat{C^*}$ to be sparser than $\mat{A}$ and $\mat{B'}$,
it reduces the overall communication volume.
The pseudocode of our algorithm is given in Section~\ref{sec:implementation}.
Like SUMMA, we operate in $\sqrt p$ rounds.
In each round, we broadcast blocks of $\mat{A^*}$ across rows of the process grid
and blocks of $\mat{B^*}$ across columns
(and we do not have to broadcast $\mat{A}$ nor $\mat{B'}$ at all).
In particular, in the $k$-th round, we broadcast block $\mat{A^*}_{k,i}$
across the $i$-th row of the process grid
(and block $\mat{B^*}_{j,k}$ across the $j$-th row),
for all $i$ (and $j$).
These broadcasts are visualized in Figure~\ref{subfig:algo-broadcast},
for matrix $\mat{A^*}$.
To make this possible, we first
need to perform one round of point-to-point communication to move
$\mat{A^*}_{k,i}$ and $\mat{B^*}_{j,k}$ to the right process row and column,
respectively. Since process $(k, i)$ communicates with its transposed rank $(i, k)$
in this round (and likewise for process $(j, k)$), each process only needs to communicate
with a single peer.
After the broadcasts are done, we compute the muliplications
$\mat{X}^i_{k,j} := \mat{A^*}_{k,i} \mat{B'}_{i,j}$
and $Y^j_{i,k} := \mat{A}_{i,j} \mat{B^*}_{j,k}$
locally on MPI process $(i, j)$.
This is done on all processes $(i, j)$ in parallel.
Finally, we compute $\mat{X}_{k,j} := \sum_{i=1}^{\sqrt p} \mat{X}^i_{k,j}$
and $\mat{Y}_{i,k} := \sum_{j=1}^{\sqrt p} \mat{Y}^j_{i,k}$
by aggregating all $\mat{X}^i_{k,j}$ and $\mat{Y}^j_{i,k}$
on process $(k,j)$ and $(i,k)$, respectively
(see Figure~\ref{subfig:algo-aggregate}).

\paragraph*{Analysis}
Like SUMMA, our algorithm
requires a communication latency of $\mathcal{O}(\sqrt p \log p)$
since we perform $\sqrt p$ rounds of collective communication
over $\sqrt p$ processes
(assuming that broadcast and aggregation steps
are implemented with a latency of $\mathcal{O}(\log p)$).
We require a communication bandwidth of
$\mathcal{O}(\max(nnz(\mat{A^*}) + nnz(\mat{B^*}), nnz(\mat{C^*}))/\sqrt p)$,
whereas SUMMA requires
$\mathcal{O}((nnz(\mat{A}) + nnz(\mat{B'}))/\sqrt p)$.
In particular, our algorithm requires less bandwidth
if $\mat{A^*}$, $\mat{B^*}$ and $\mat{C^*}$ are considerably sparser
than $\mat{A}$ and $\mat{B'}$.
Let $\mathrm{flops}$ denote the number of scalar multiplications
required to form $\mat{C}^*$. SUMMA
requires $\mathcal{O}(\mathrm{flops}/p)$
time for local computation,
while our algorithm requires $\mathcal{O}((\mathrm{flops}\log p)/p)$
(assuming that a $(\log p)$-round parallel reduction is used for aggregation).
Our algorithm consumes at most
$\mathcal{O}(\max(nnz(\mat{A'}), nnz(\mat{B'}), nnz(\mat{C'}))/p)$
memory.


\subsection{Algorithm for General Updates}
\label{subsec:general-spgemm}

\begin{algorithm}[t]
\caption{MPI-parallel dynamic SpGEMM, general updates}
\label{algo:general-dynamic}
\begin{algorithmic}
\State $\mat{C^*}, \mat{F^*} \gets \Call{computePattern}{\mat{A}, \mat{A^*}, \mat{B}, \mat{B^*}}$
\State $\mat{E}_{i,j} \gets \mat{F}_{i,j} \oplus \mat{F^*}_{i,j}$
	masked at $\mat{C^*}_{i,j}$ \Comment{local}
\State aggregate($\oplus$) $\mat{E}_{i,j}$ to $\vec{R_i}$ over $j$-th process row
\State $\mat{A}^\vec{R}_{i,j} \gets \mat{A'}$ filtered by $\vec{R_i}$ \Comment{local}
\State send $\mat{A}^\vec{R}_{i,j}$ to process $(j,i)$, receive $\mat{A}^\vec{R}_{j,i}$
\For{$k = 1, \ldots, \sqrt p$}
	\State broadcast $\mat{A}^\vec{R}_{k,i}$ over $i$-th process row
	\State broadcast $\mat{C^*}_{k,j}$ over $j$-th process column
	\State $\mat{Z}^i_{k,j}, \mat{H}^i_{k,j} \gets \mat{A}^\vec{R}_{k,i} \mat{B'}_{i,j}$
		masked at $\mat{C^*}_{k,j}$ \\\Comment{local, returns the new Bloom filter in $\mat{H}^i_{k,j}$}
	\State aggregate $\mat{Z}^i_{k,j}$ to $\mat{Z}_{k,j}$ over $j$-th process column
	\State aggregate($\oplus$) $\mat{H}^i_{k,j}$ to $\mat{H}_{k,j}$ over $j$-th process column
\EndFor
\State merge $\mat{Z}_{i,j}$ into $\mat{C}_{i,j}$, masked at $\mat{C^*}_{i,j}$, to obtain $\mat{C'}_{i,j}$
	\Comment{local}
\State merge $\mat{H}_{i,j}$ into $\mat{F}_{i,j}$, masked at $\mat{C^*}_{i,j}$, to obtain $\mat{F'}_{i,j}$
	\Comment{local}
\end{algorithmic}
\end{algorithm}

For general updates, it is not enough to communicate the update matrices
among MPI processes. Since general updates are not restricted in any way,
computing the result of the dynamic matrix multiplication can require arbitrary
entries of the new input matrices $\mat{A'}$ and $\mat{B'}$
(including entries that were \emph{not} changed in comparison to $\mat{A}$ and $\mat{B}$).
Fortunately, we can still reduce the communication volume and computational cost
compared to a static recomputation since not all entries of $\mat{C'}$
need to be recomputed.
Our approach is to use a masked SpGEMM for this purpose
that we specialize for the problem of dynamically updating $\mat{C'}$.
For any masked SpGEMM, it is straightforward to identify the rows of $\mat{A'}$
and the columns of $\mat{B'}$ that can contribute
to a given set of entries in $\mat{C'}$ (by just considering
the appropriate rows/columns that have non-zero entries in the mask).
However, our algorithm can further restrict the data that we need
to exchange by considering only some \emph{columns} of $\mat{A'}$
and some \emph{rows} of $\mat{B'}$.
For this purpose, we use a Bloom filter that remembers which
of the terms $a_{i,k} b_{k,j}$
contribute to any $c_{i,j}$.\footnote{A Bloom filter was previously
	used by Azad \etal~\cite{DBLP:conf/ipps/AzadBG15} for (non-dynamic) masked SpGEMM.
	However, while their approach
	uses a Bloom filter to exclude rows of the left-hand side that
	need to be communicated, we use the exact mask (and not the Bloom filter)
	to avoid communicating rows of the left-hand side,
	and use the Bloom filter to additionally exclude columns.}
More specifically, our Bloom filter is a matrix $\mat{F}$
that stores an $\ell$-bit bitfield in
each entry $f_{i,j}$, where $\ell$ is a constant (in practice, we use $\ell = 64$).
While computing $\mat{C} = \mat{A} \mat{B}$, we set the $(k \mod \ell)$-th bit
of $f_{i,j}$ to 1 if the term $a_{i,k} b_{k,j}$ contributes to the
value of $c_{i,j}$ (and $f_{i,j} = 0$ if there is no such $k$).
Given this bitmask, we can later recover a superset of the columns
of $\mat{A}$ (or rows of $\mat{B}$) that contributed to
$c_{i,j}$.

Our algorithm is given in Algorithm~\ref{algo:general-dynamic}.
We first compute $\mat{C^*}$ as given in Eq.~(\ref{eq:cupdate}).
This computation is done using the algorithm for algebraic updates.
We modify that algorithm to also compute a Bloom filter $\mat{F^*}$
such that bit $(k \mod \ell)$ is set in $f^*_{i,j}$ whenever
the terms $a^*_{i,k} b'_{k,j}$ or $a'_{i,k} b^*_{k,j}$ contribute
to $c^*_{i,j}$. This computation is denoted by \textsc{computePattern}
in our pseudocode. We remark that we do not require the values of
$\mat{C^*}$ for our algorithm; computing the sparsity structure
of $\mat{C^*}$ is enough.

Given $\mat{F^*}$, we compute the matrix $\mat{F} \oplus \mat{F^*}$,
where $\oplus$ denotes bitwise or.
We form the matrix $\mat{E}$ by keeping only the
entries of $\mat{F} \oplus \mat{F^*}$ that are non-zero in $\mat{C^*}$.
This matrix acts as a Bloom filter that can be used to select
a superset of the columns of $\mat{A'}$ and rows of $\mat{B'}$
that are needed to compute $\mat{C'}$; considering $\mat{F^*}$
is needed to account for columns/rows that are required
due to new non-zeros in $\mat{A'}$ and $\mat{B'}$
(compared to $\mat{A}$ and $\mat{B}$).
We reduce $\mat{E}$ over its rows (via bitwise or).
The result is a vector $\vec{R}$ such that bit
$(k \mod \ell)$ is set in $r_i$ if there is any column $j$ of $\mat{C'}$
such that $a'_{i,k} b'_{k,j}$ is required to compute $c'_{i,j}$.
We now extract the rows $i$ of $\mat{A'}$ such that $r_i$ is non-zero
and only extract the columns $k$ such that bit $(k \mod \ell)$
is set in $r_i$. This yields the matrix $\mat{A}^\vec{R}$.
While it would also be possible to filter
(and broadcast, in the next step of the algorithm)
$\mat{B'}$ instead of $\mat{A'}$, we chose $\mat{A'}$
because our matrices are locally stored row-wise.
Hence, we can efficiently extract specific rows of $\mat{A}$
and then discard a subset of the columns from these rows.

The remainder of the algorithm proceeds similarly to the
algorithm for the algebraic case. We broadcast
$\mat{A}^\vec{R}$ over the rows of the process grid
(similarly to the algebraic case, see Figure~\ref{subfig:algo-broadcast}).
To be able to make use of $\mat{C^*}$ as an output mask
during the local multiplication,
we broadcast this matrix over the columns of the process grid
and perform a local masked matrix multiplication.
This local multiplication also produces an updated
Bloom filter (called $\mat{H}^i_{k,j}$ in the pseudocode).
Finally, we aggregate both the updated entries
of the matrix (called $\mat{Z}^i_{k,j}$ in Algorithm~\ref{algo:general-dynamic})
and updated entries of the Bloom filter onto MPI process $(i, j)$.

\paragraph*{Analyis}
Following a similar analysis as in the algebraic case,
we find that our communication latency is $\mathcal{O}(\sqrt p \log p)$
(which is identical to the SUMMA algorithm).
The communication bandwidth is dominated by the broadcasts,
\ie $\mathcal{O}((nnz(\mat{A}^\vec{R}) + nnz(\mat{C^*}))/\sqrt p)$.
Local computation costs are $\mathcal{O}(\mathrm{flops} \log p)/p$,
where $\mathrm{flops}$ is the number of scalar multiplications
required to form $\mat{A}^\vec{R} \mat{B'}$.
While $nnz(\mat{A}^\vec{R}) = nnz(\mat{A'})$ in the worst case,
we expect that our Bloom filter allows us to discard
many non-zeros of $\mat{A'}$ without considering them in the
computation.


\subsection{Handling Transposition}
\label{sec:algo-transpose}

Given an efficient local algorithm for SpGEMM that supports transposition,
our algorithms can naturally be extended to the case where
$\mat{A}$ and/or $\mat{B}$ are transposed.
In the algorithm for algebraic updates, we can simply replace $\mat{A^*}_{k,i}$
and $\mat{B^*}_{j,k}$ by $\mat{A^*}_{i,k}$ and/or $\mat{B^*}_{k,j}$
if $\mat{A}$ and/or $\mat{B}$ are transposed.
Furthermore, if $\mat{A}$ and/or $\mat{B}$ are transposed,
we need to broadcast $\mat{B^*}$
over rows and/or $\mat{A^*}$ over columns of the process grid.
In some cases, this allows us to get rid of the
initial send/receive call
since the blocks that are broadcasted are already on the right
process row and/or column.
Finally, the local matrix multiplication
algorithm has to take transposition into account.

A similar strategy can be applied in the case of general updates.
We note that in the transposed case, the Bloom filter
can be used to discard rows of $\mat{A'}$ and columns
of $\mat{B'}$, respectively. Hence, it is still possible to
filter the matrices efficiently, even if transposition is applied.


\section{Implementation Details}
\label{sec:implementation}

\subsection{Algebraic SpGEMM}

Our implementation stores $\mat{C'}$ as a dynamic matrix.
$\mat{A^*}$ and $\mat{B^*}$ are stored in DCSR format as we expect them
to be hypersparse. We also use the DCSR format when broadcasting blocks
of matrices.

Our local multiplication uses Gustavson's row-wise sparse
matrix multiplication algorithm~\cite{10.1145/355791.355796}.
We use shared-memory parallelism to parallelize the computation
of different rows of the result. Each shared-memory thread uses
a sparse accumulator based on a dynamic array combined with a hash table
for this purpose. We concatenate all output rows into a DCSR
to form $\mat{X}^i_{k,j}$ and $\mat{Y}^j_{i,k}$.

Since $\mat{X}^i_{k,j}$ and $\mat{Y}^j_{i,k}$ are expected to have
different sparsity patterns, we cannot use a straightforward
MPI \textsc{Reduce} call to aggregate them.
Instead, we use an approach based on a custom reduce-scatter
implementation for sparse matrices.
Since the output of that aggregation (\ie $\mat{C'}$) is a dynamic matrix
that supports efficient local updates, we do not need any auxiliary
data structure (such as a SPA~\cite{Gilbert92sparsematrices} or similar) during aggregation.

\subsection{General SpGEMM}

As in our algorithm for algebraic updates, we store
the result matrices $\mat{C'}$ and $\mat{F'}$ as dynamic matrices
and all intermediate matrices as DCSR.

To efficiently compute the local masked matrix multiplication,
we locally build a hash table that stores the indices $(u, v)$
of all non-zeros in $\mat{C^*}_{k,j}$ on process $(i, j)$.
While this duplicates the work of building the same hash table
on multiple MPI processes,
we found it to be faster than broadcasting the hash table itself
(instead of $\mat{C^*}_{k,j}$) in preliminary experiments;
this is caused by the fact that the hash table is considerably
larger than $nnz(\mat{C^*}_{k,j})$ due to its empty slots.
We then perform a variant of Gustavson's row-wise sparse
matrix multiplication that uses the hash table to check whether
$(u, v)$ is non-zero in $\mat{C^*}$ before adding an index pair
$(u, v)$ to the sparse aggregator.


\section{Experiments}
\label{sec:experiments}
In this section, we present experiments to evaluate the performance of our algorithms in practice.
We have implemented our algorithms in
C++. The code of our algorithms will be published as open source
	software once this paper is accepted.
\change{We use CombBLAS 2.0, CTF 1.35 and PETSc 3.17.1 as state-of-the-art competitors.
We note that CombBLAS seems to outperform both CTF and PETSc on our benchmarks;
this result is in line with results by the authors of CombBLAS~\cite{9470983}.}


\subsection{Experimental Setup}
\label{app:experiment-setup}

\begin{table}[tb]
	\centering
	\caption{List of real-world instances.}
	\label{table:instances}
	\begin{tabular}{lllrr}
		\toprule
		Instance & Source & Type & $n$ & $nnz$ \\
		\midrule
		LiveJournal & SNAP & Social & $\numprint[M]{4}$ & $\numprint[M]{86}$ \\
orkut & SNAP & Social & $\numprint[M]{3}$ & $\numprint[M]{234}$ \\
tech-p2p & Network Repository & Peer-to-Peer & $\numprint[M]{5}$ & $\numprint[M]{295}$ \\
indochina & Network Repository & Web & $\numprint[M]{7}$ & $\numprint[M]{304}$ \\
sinaweibo & Network Repository & Social & $\numprint[M]{58}$ & $\numprint[M]{522}$ \\
uk2002 & Network Repository & Web & $\numprint[M]{18}$ & $\numprint[M]{529}$ \\
wikipedia & Network Repository & Web & $\numprint[M]{27}$ & $\numprint[M]{1088}$ \\
PayDomain & Network Repository & Web & $\numprint[M]{42}$ & $\numprint[M]{1165}$ \\
uk2005 & Network Repository & Web & $\numprint[M]{39}$ & $\numprint[M]{1581}$ \\
webbase & Network Repository & Web & $\numprint[M]{118}$ & $\numprint[M]{1736}$ \\
twitter & Network Repository & Social & $\numprint[M]{41}$ & $\numprint[M]{2405}$ \\
friendster & SNAP & Social & $\numprint[M]{124}$ & $\numprint[M]{3612}$
\\
		\bottomrule
	\end{tabular}
\end{table}

The experiments are performed on a 16-node compute cluster. Each compute
node of the cluster features two Intel Xeon 6126 CPUs with 12 cores per CPU, and 192 GB RAM.
The cluster is connected using
100 GBit Intel Omni-Path Architecture interconnects.
For CombBLAS, CTF and our algorithms, we run 4 MPI processes per node
(\ie two MPI processes per CPU socket) as these frameworks require a square processor grid.
In this configuration, we run 6 OpenMP threads per MPI process.
\change{For PETSc, we found that a single MPI process per node
(and 24 threads per MPI process) yields the best
performance.\footnote{In experiments that have a fixed input size per
	MPI process (\eg weak scaling experiments below),
	we adjust the input size of PETSc
	to ensure that all competitors operate on the same
	number of non-zeros.}}
Except in the parallel scalability experiments, we use all 16 compute nodes
in each experiment.

Our experimental data consists of several large graphs -- ranging from
86 million to almost 4 billion edges --
and is shown in Table~\ref{table:instances}. All instances were downloaded from the
SNAP datasets~\cite{snapnets} and Network Repository~\cite{networkrepository}.
We always read graphs as undirected when creating adjacency matrices,
\ie for an edge $\{u, v\}$ in the input data, we add non-zeros
$(u, v)$ and $(v, u)$ to the matrix.

The instances we use demonstrate significant imbalance without remapping.
To avoid load imbalance, we randomly permute input indices before
constructing each matrix.
Given our 2D representation (and the memory layouts of our competitors),
random permutation provides an adequate distribution
of the input data.
The same I/O and same permutation method is used for our code and for
our competitors.
We do not measure I/O times in any of our experiments.


%

\subsection{Performance of Data Structures for Dynamic Distributed Graphs}
\label{app:experiments-ops}
\begin{figure}[tb]
	\centering
	\includegraphics[scale=.5]{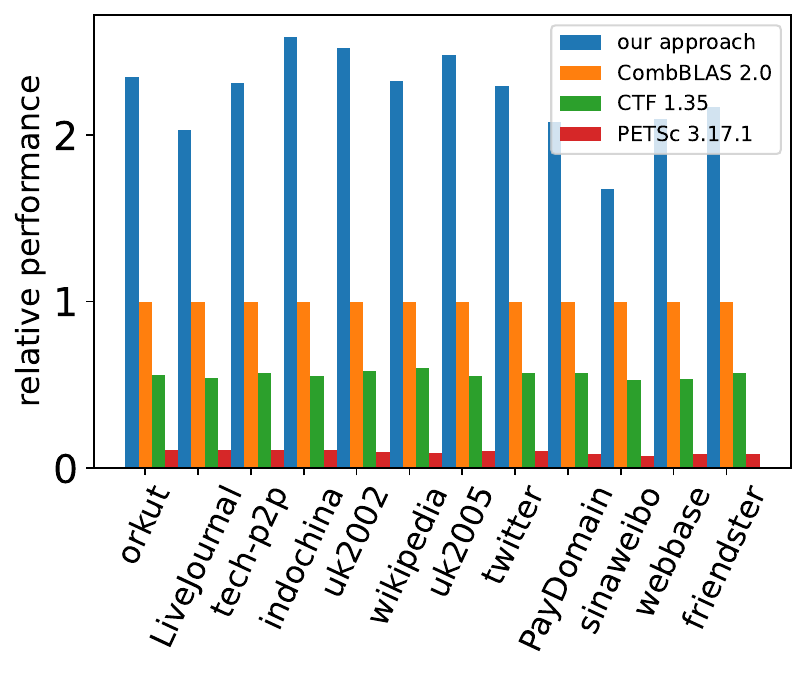}
	\caption{Construction}
	\label{subfig:construction} 
	\caption{\change{Matrix construction performance
		on real-world graphs. Performance is measured relative
		to CombBLAS.}}
\end{figure}
\begin{figure}[tb]
	\centering
	\includegraphics[scale=.35]{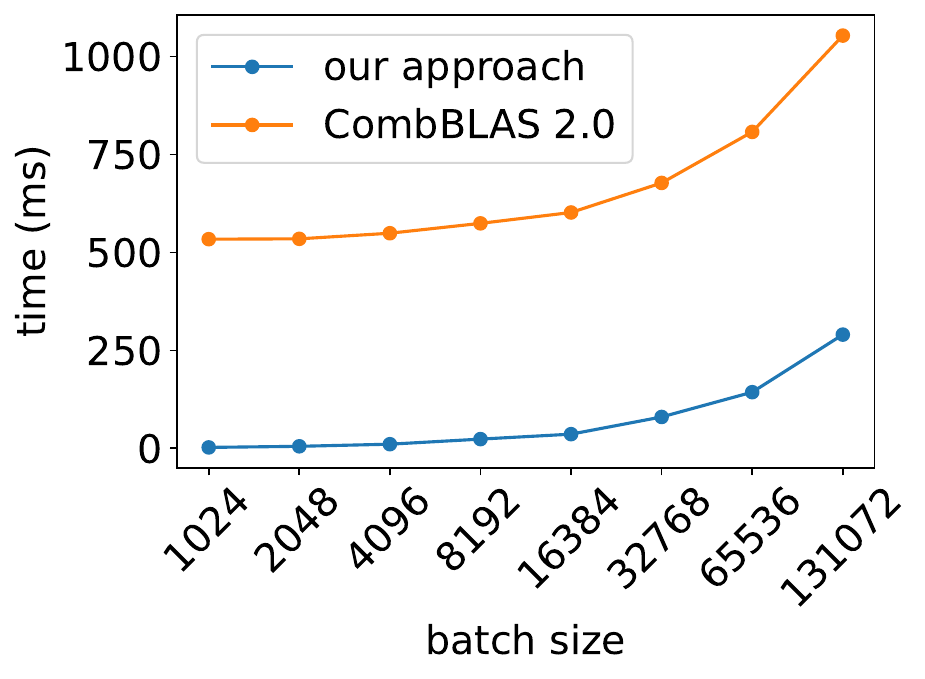}
	\caption{\change{Mean insertion performance
		on real-world graphs. CTF and PETSc
		are considerably slower than our algorithm
		(at least $\insertionctfminspeedup \times$ slower,
		and $\insertionpetscminspeedup \times$ slower, respectively)
		and not plotted.}}
	\label{subfig:insertion}
\end{figure}
\begin{figure}[tb]
	\begin{subfigure}[t]{0.45\columnwidth}
		\centering
		\includegraphics[scale=.28]{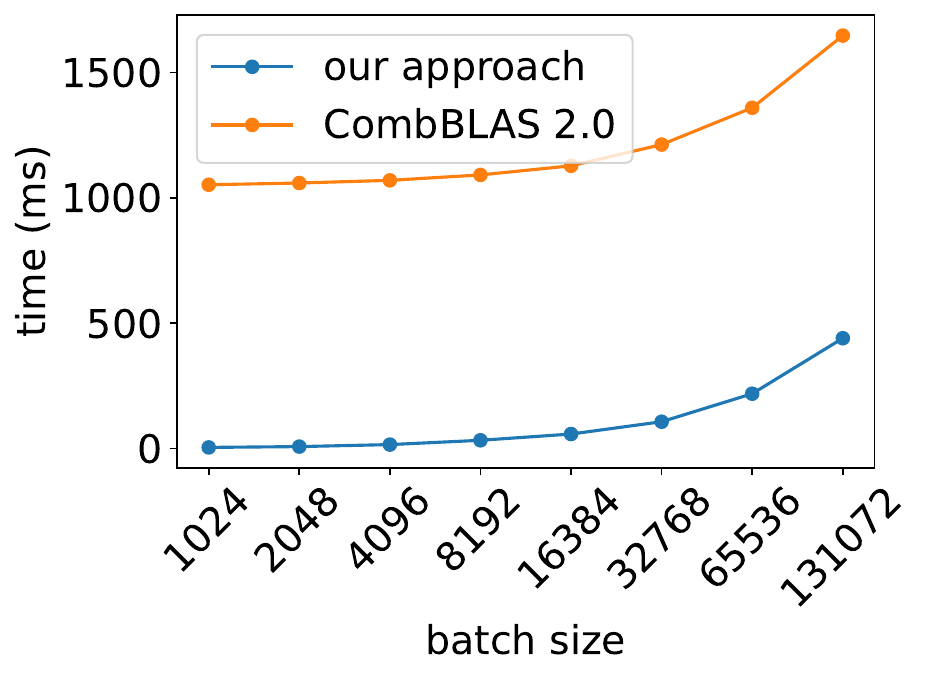}
		\caption{Updates}
		\label{subfig:updates}  
	\end{subfigure}\hfill%
	\begin{subfigure}[t]{0.45\columnwidth}
		\centering
		\includegraphics[scale=.28]{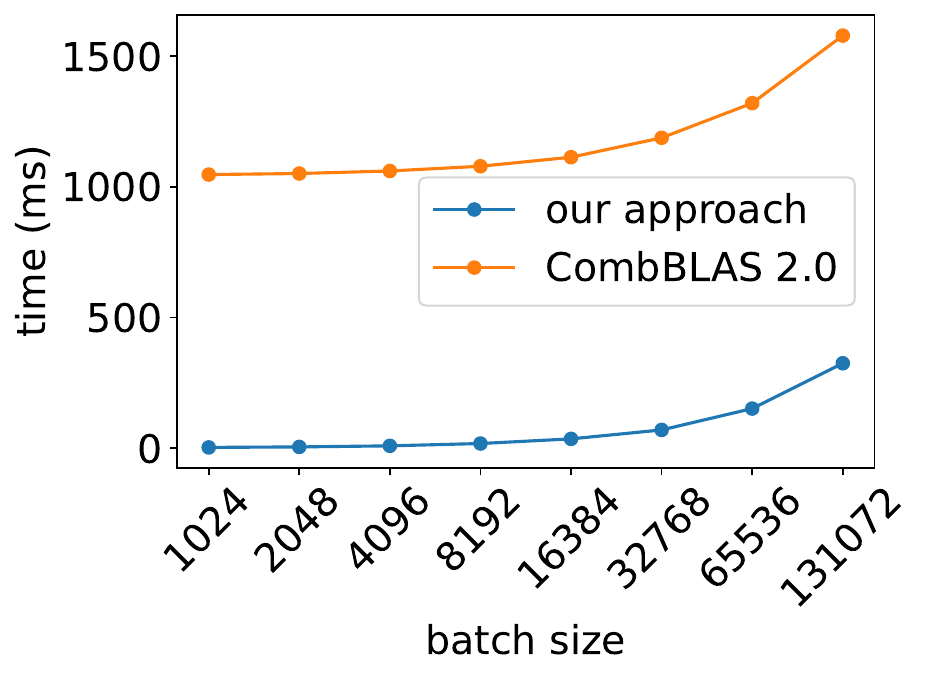}
		\caption{Deletion}
		\label{subfig:deletion}
	\end{subfigure}
	\caption{\change{Mean update and deletion performance
		on real-world graphs. CTF and PETSc
		are not plotted, but at least
		$\updatectfminspeedup \times$
		and $\updatepetscminspeedup \times$ slower than our algorithm
		for updates.
		For deletions, CTF is at least $\deletionctfminspeedup \times$
		slower than our algorithm (PETSc does not support efficient deletions).}}
\end{figure}
\begin{figure}[tb]
	\centering
	\includegraphics[scale=.35]{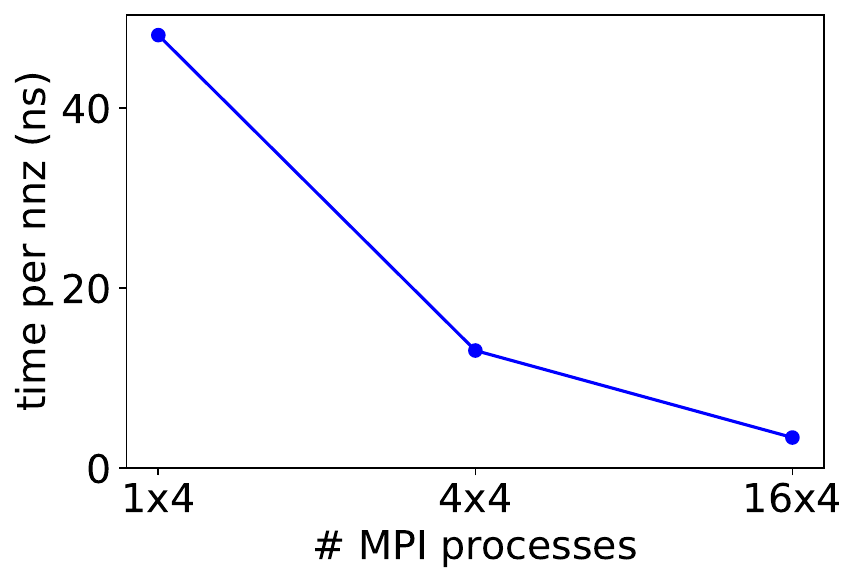}
	\caption{\change{Weak scalability of insertions on real-world graphs,
		for different numbers of compute nodes. Each
		MPI process performs 1.3M insertions.}}
	\label{subfig:insert-scale}
\end{figure}
\begin{figure}[tb]
	\centering
	\includegraphics[scale=.35]{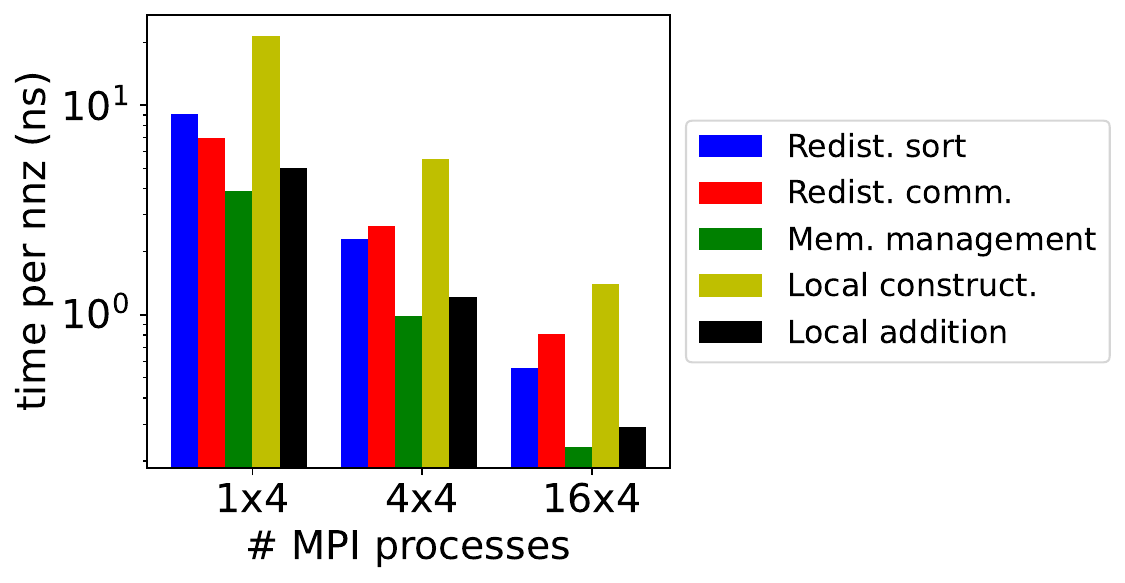}
	\caption{\change{Breakdown of insertion running time on real-world graphs,
		for different numbers of compute nodes.}}
	\label{subfig:insert-breakdown}
\end{figure}%

In our first experiments, we evaluate the performance
and parallel scalability of our dynamic distributed matrix data
structure and our redistribution algorithm for updates.

\paragraph{Construction} 
We measure the time that it takes to
construct the adjacency matrix of each input graph.
We use our dynamic matrix data structure for this purpose;
for CombBLAS, we use its default DCSC data structure, PETSc uses
its parallel compressed sparse row implementation.

Figure~\ref{subfig:construction} depicts the results of
this experiment. We report the performance
relative to CombBLAS.
Our redistribution algorithm outperforms all state-of-the-art competitors.
Our code is between $\constructioncombblasminspeedup \times$
and $\constructioncombblasmaxspeedup \times$ faster than CombBLAS
(the best competitor); both CTF and PETSc are slower
than both CombBLAS and our code on every instance.
This difference in performance is due to two reasons:
first, our dynamic matrix data structure
allows for fast insertion and exploits shared-memory parallelism
effectively.
Secondly, our redistribution of non-zeros is faster than
the competitor's
(which consists of a comparison sort and a global \textsc{AllToAll}
in the case of CombBLAS).
In fact, even if we construct a DCSR
(\ie the same type of data structure that CombBLAS uses)
instead of a dynamic matrix, we are still
on average $\constructiondcsrspeedup \times$
faster than CombBLAS, although CombBLAS constructs its DCSC
data structure faster on some instances.

\paragraph{Insertions}
We now compare the performance
of inserting new non-zeros into the adjacency matrix,
for our implementation versus the state-of-the-art competitors.
In this experiment, we insert half of the non-zeros
initially; this does not contribute to
the running time we measure. Afterwards,
we insert randomly chosen non-zeros from the remaining
half into the already existing matrix,
by first constructing an update matrix
(in hypersparse layout for the competitors that support it)
and adding the update matrix to the adjacency matrix
(which uses a dynamic storage for our implementation).
Insertions are performed in batches of various sizes.
The batch size denotes the number of insertions
that \emph{each} MPI process performs
-- \ie the full update matrix has (batch size $\times$ $p$)-many non-zeros.
In our experiments, this results in between 65K to 8.3M non-zeros in each update matrix.
We perform 10 batches per instance (such that up to 83M entries are inserted
into the already existing matrix).

Results are depicted in Figure~\ref{subfig:insertion}.
Since all competitors have to use a static matrix data structure that
they have to rebuild
after insertions, we outperform the competitors.
In particular, we outperform
CombBLAS by
$\insertioncombblasminspeedup \times$
(for batch size 131072) to $\insertioncombblasmaxspeedup \times$
(for batch size 1024).
As expected, our speedup over CombBLAS decreases with
the batch size; as update matrices become denser, the cost
of rebuilding the output matrix amortizes more effectively.
Compared to CTF, we are always at least $\insertionctfminspeedup \times$
faster, and compared to PETSc, we are at least $\insertionpetscminspeedup \times$
faster.

\paragraph{Updates and Deletions}

To evaluate updates and deletions of the matrix,
we proceed similarly to the insertion experiment.
However, for update and deletion experiments, we
insert the full adjacency matrix initially
(and only draw non-zeros for the update matrix
from existing non-zeros of the adjacency matrix).
We note that PETSc does not support an efficient way
to mask non-zeros in matrices; thus, we do not compare
against PETSc for deletions.

The results are visualized in Figure~\ref{subfig:updates} and
~\ref{subfig:deletion}.
Similarly to the insertion case, we are $\updatecombblasminspeedup \times$ to
$\updatecombblasmaxspeedup \times$ faster for updates and
$\deletioncombblasminspeedup \times$ to $\deletioncombblasmaxspeedup \times$
faster for deletions than CombBLAS.
Likewise, our algorithm performs updates at least $\updatectfminspeedup \times$
faster than CTF and at least $\updatepetscminspeedup \times$ faster than PETSc.
For deletions, we are always at least $\deletionctfminspeedup \times$ faster than CTF.


\paragraph{Parallel Scalability of Graph Updates}
\begin{figure}
	\begin{subfigure}[t]{.49\columnwidth}
		\centering
		\includegraphics[scale=.28]{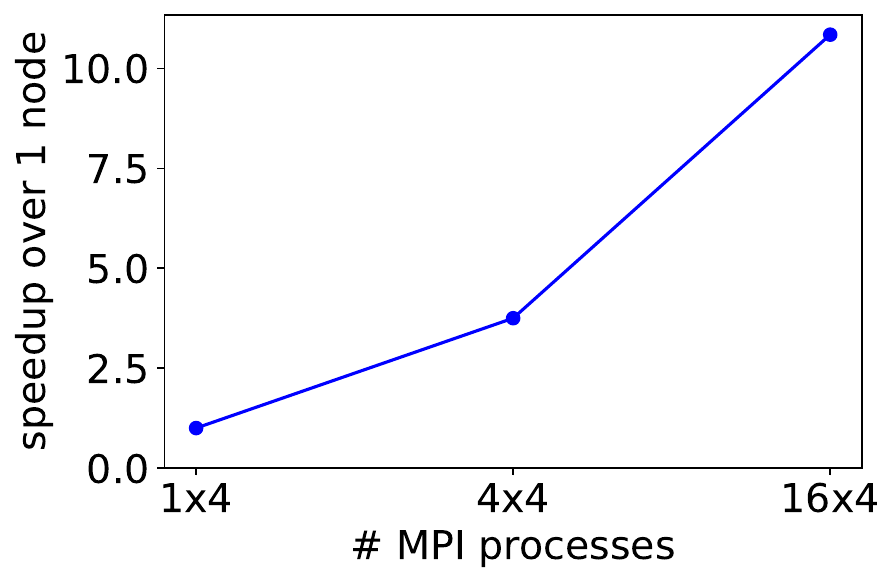}
		\caption{Strong scalability, $2^{30}$ insertions in total.}
	\end{subfigure}\hfill%
	\begin{subfigure}[t]{.49\columnwidth}
		\centering
		\includegraphics[scale=.28]{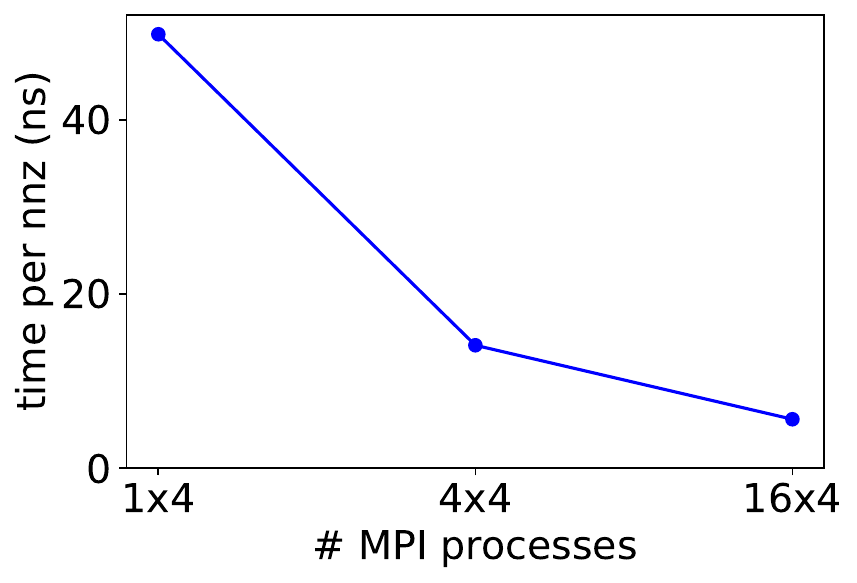}
		\caption{\change{Weak scalability, $2^{28}$ insertions per MPI process.}}
	\end{subfigure}
	\caption{Parallel scalability of insertions on synthetic R-MAT graphs,
		for different numbers of compute nodes.}
	\label{fig:insertion-scalability-synthetic}
\end{figure}

We evaluate the scalability of insertions across
different numbers of compute nodes.
We vary the number of compute nodes between 1, 4 and 16
(such that the process grid is square), while
keeping the OpenMP thread count constant at 6,
and the number of MPI processes per node constant at 4.
Insertions are performed in the same way as in the previous section,
with the batch size fixed to 131072.
This can be seen as a weak scaling experiment since $\mathrm{nnz}/p$
remains constant, where $\mathrm{nnz}$ denotes the number
of non-zeros in the update matrix.

Figure~\ref{subfig:insert-scale} depicts the results
of this experiment. In particular, the time per non-zero
decreases with increasing numbers of compute nodes,
indicating that our algorithm does not yet hit
a scalability bottleneck on our cluster.
%
%
Figure~\ref{subfig:insert-breakdown}
shows that all steps of our algorithm
scale well with the number of compute nodes.
A large fraction of the running time consists of
local operations as opposed to communication.

\paragraph{Scalability on Synthetic Graphs}
In Figure~\ref{fig:insertion-scalability-synthetic}, we present
strong scaling and weak scaling results for insertions into our
data structure on synthetic R-MAT graphs.
We use the same R-MAT parameters as the Graph500 benchmark.
In the strong scaling experiment, each MPI process generates
$2^{30}/p$ non-zeros according to the R-MAT model.
In the weak scaling experiment, each process generates $2^{28}$ non-zeros.
As in our experiments on real-world graphs, we apply
a (global) permutation to the row/column indices to ensure
that load is evenly balanced across the process grid.
We use a batch size of 131072 entries
and insert all entries into a dynamic matrix.
Our algorithm scales well with
increasing numbers of MPI processes: for 16 compute nodes, we
achieve a strong speedup of $\constructionsynthstrongspeedup \times$
over a single compute node.
Likewise, in the weak scaling model, our time per non-zero
drastically decreases with the number of compute nodes,
indicating that our algorithm does not hit an efficiency
bottleneck in the configurations that we tested.

\subsection{Performance of Dynamic SpGEMM}

We perform experiments on both dynamic SpGEMM algorithms that are
presented in Section~\ref{sec:dynamic-distributed-spgemm}.

\begin{figure}
	\centering
	\includegraphics[scale=.5]{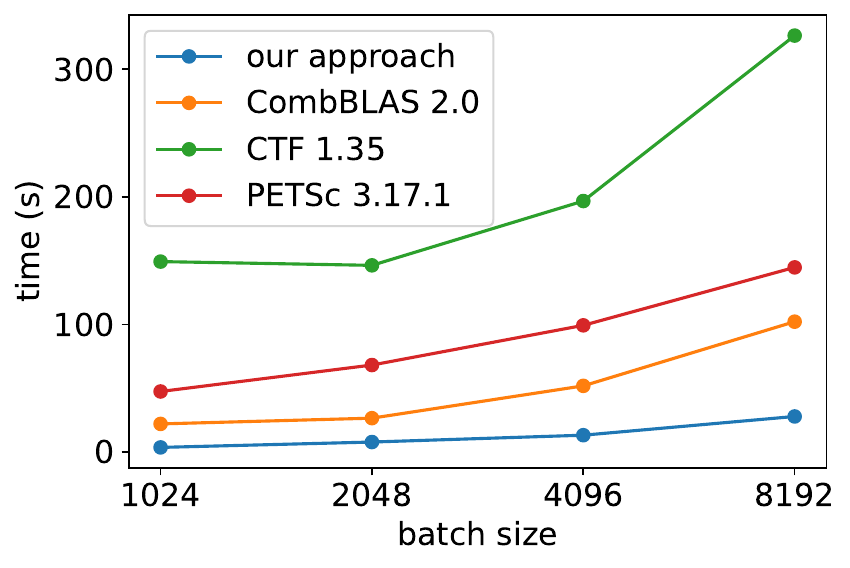}
	\caption{Mean performance of dynamic SpGEMM, \textbf{algebraic case},
		on real-world graphs.
		Batch sizes
		are \emph{per MPI process}, \ie update matrices have up to 524K non-zeros.}
	\label{subfig:spgemm-combblas}
\end{figure}
\begin{figure}
	\centering
	\includegraphics[scale=.5]{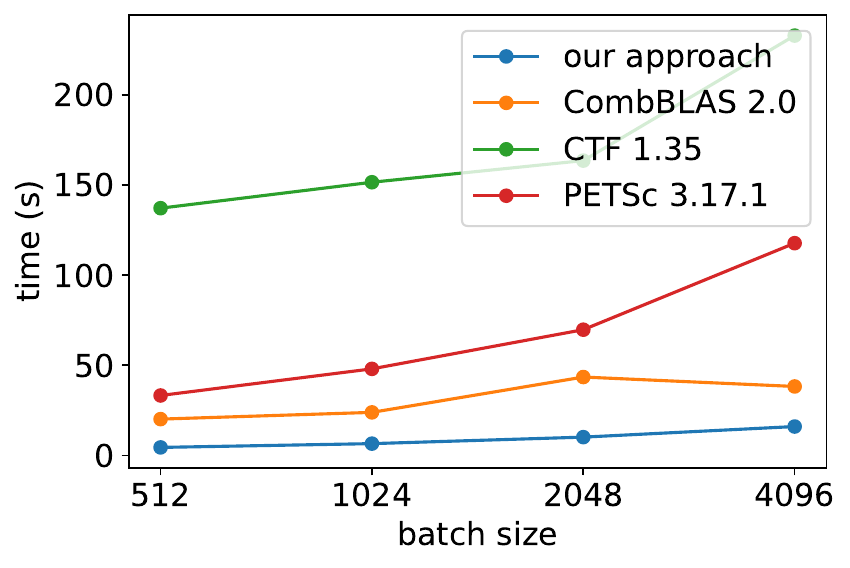}
	\caption{Mean performance of dynamic SpGEMM, \textbf{general case},
		on real-world graphs.
		Batch sizes
		are \emph{per MPI process}, \ie update matrices have up to 262K non-zeros.}
	\label{subfig:general-spgemm-combblas}
\end{figure}


\paragraph{Algebraic case}
This experiment repeatedly computes $\mat{C'} = \mat{A'}\mat{B}$, where
in each iteration we insert non-zeros into $\mat{A'}$, while $\mat{B}$ is static.
\change{In particular, we initialize $\mat{A'}$ to an empty matrix (\ie the zero matrix)
for each group.
$\mat{B}$ is initialized to the full adjacency matrix of the graph.
Initialization of $\mat{B}$ is not included
in the measured runtime.} Afterwards, insertions are
performed in batches, with \change{batch sizes between 1024 and 8192 vertices
per MPI process}. These insertions
equate to the expression $\mat{A'} = \mat{A} + \mat{A^*}$, where $\mat{A^*}$
is the matrix of insertions.
We draw the insertions from the adjacency matrix of the graph
(\ie drawing all possible insertions would result in the computation of $\mat{A}^2$).
\change{Each MPI process draws insertions individually, independently,
and uniformly at random. Furthermore, the method (and random seed) to draw non-zeros is the same
for our competitors and for our approach.}
We perform 10 batches per instance, therefore
the total number of non-zeros in the left-hand side varies between 655K and 5.2M
\change{(= number of batches per instance $\times$ batch size $\times$ number of MPI processes)}.
Batch sizes
of over 8192 produced result matrices that did not fit in the RAM of our compute
cluster, both for the competitors and for our algorithm, hence they were excluded.
We perform the multiplication in the $(+, \cdot)$ semiring,
so we can utilize the algorithm for algebraic updates
presented in Section~\ref{sec:spgemm-algebraic}.
As previously demonstrated, our algorithm reduces $\mat{C'} = \mat{A'}\mat{B}$
to $\mat{C'} = \mat{C} + \mat{A^*}\mat{B}$, provided $\mat{B}$ is static.
Our competitors compute $\mat{A^*}\mat{B}$ using their distributed SpGEMM algorithms
and add the result to $\mat{C}$.

The experiment (results shown in Figure~\ref{subfig:spgemm-combblas}) demonstrates that our algorithm is
$\dynamicspgemmcombblasminspeedup \times$ (for a batch size of 8192)
to $\dynamicspgemmcombblasmaxspeedup \times$ (for batch size 1024) faster than CombBLAS
(which is the best competitor in this experiment).
We are also at least $\dynamicspgemmctfminspeedup \times$ faster than CTF
and at least $\dynamicspgemmpetscminspeedup \times$ faster than PETSc.
As expected, the speedup decreases for increasing batch sizes.
In particular, for large batch sizes, update matrices are not hypersparse
on all instances anymore.
In these cases, our algorithm is expected to perform worse than SUMMA due to its
more complicated communication pattern.

\paragraph{General case}
This experiment is performed using the same setup as before, but using the general dynamic
SpGEMM algorithm described in Section~\ref{subsec:general-spgemm} instead. We use a $(\min, +)$ semiring
to differentiate from the algebraic case.
To perform an equivalent operation using our competitors, it is no longer enough to compute
just $\mat{A^*}\mat{B}$, we have to recompute $\mat{A'}\mat{B}$ from scratch
(since insertions into the matrix are incompatible with the $\min$ operator
that the semiring uses for addition).
\change{Unlike our other competitors,
PETSc does not support arbitrary semirings; thus, we continue to use the
$(+, \cdot)$ semiring for PETSc in this experiment.}

The results
for this experiment are depicted in Figure~\ref{subfig:general-spgemm-combblas}. Our findings
show that -- depending on the batch size -- our algorithm
is $\generalspgemmcombblasminspeedup \times$ to $\generalspgemmcombblasmaxspeedup \times$ faster
than CombBLAS, the best competitor in this experiment.
Additionally, we are always at least $\generalspgemmctfminspeedup \times$ faster than CTF
and at least $\generalspgemmpetscminspeedup \times$ faster than PETSc.
While our competitors communicate all non-zeros of the left-hand side, our algorithm
can avoid this using our Bloom filter for entries that do not contribute to the result. As the matrix becomes denser,
the probability that a non-zero does not contribute to the result decreases. Hence
for larger batch sizes it is expected to be more efficient to simply transfer all non-zeros, because
the overhead of a Bloom filter is less economical.

\begin{figure}[tb]
	\centering
	\includegraphics[scale=.35]{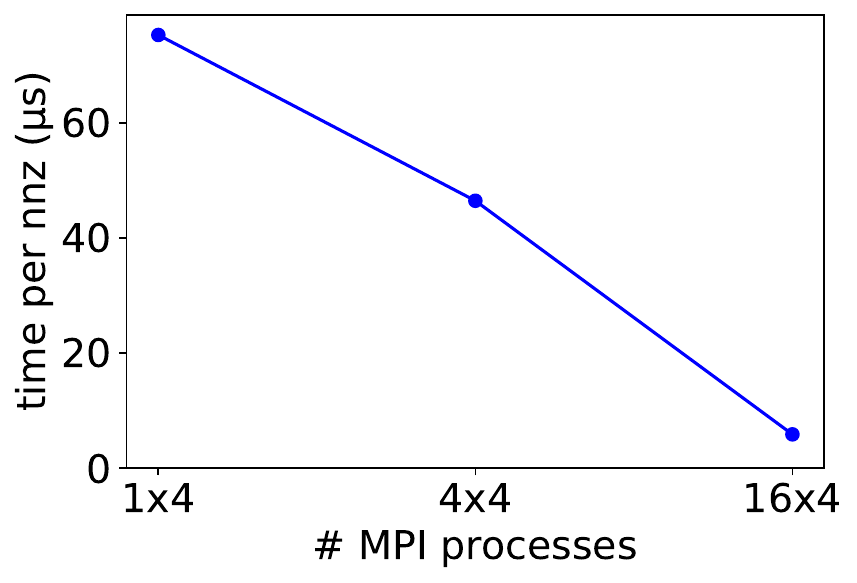}
	\caption{\change{Weak scalability of dynamic SpGEMM (algebraic case)
		on real-world graphs, for different numbers of compute nodes.
		This experiment uses 81920 non-zeros per MPI process.}}
	\label{subfig:spgemm-scale}
\end{figure}
\begin{figure}[tb]
	\centering
	\includegraphics[scale=.35]{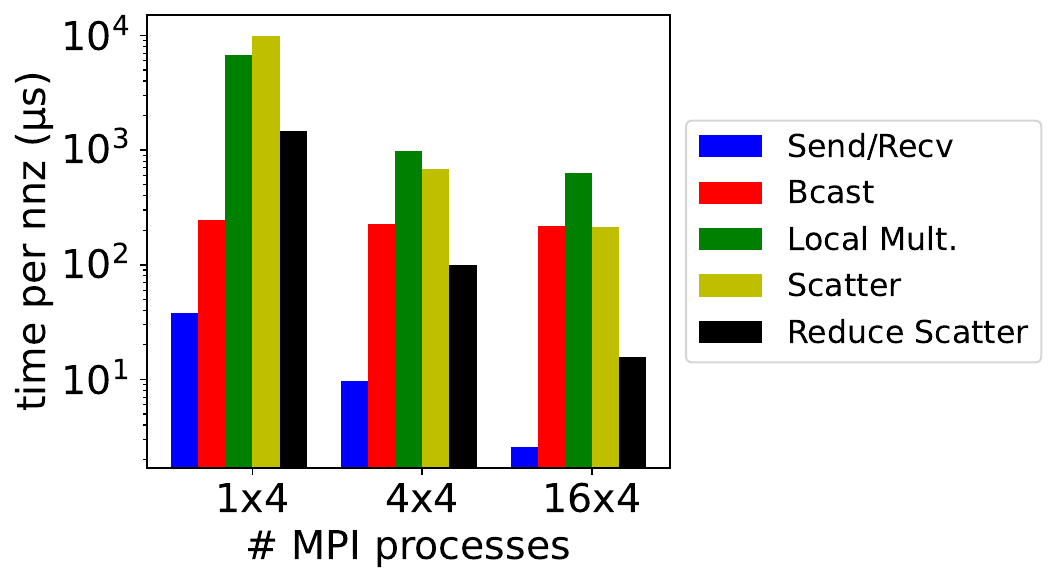}
	\caption{\change{Breakdown of dynamic SpGEMM (algebraic case) running time
		on real-world graphs,
		for different numbers of compute nodes.}}
	\label{subfig:spgemm-breakdown}
\end{figure}

\paragraph{Parallel Scalability of Dynamic SpGEMM}

To test the scalability of our algorithm, we perform the SpGEMM experiment
with algebraic updates
on a varying number of MPI processes.
We change the number of compute nodes between 1, 4 and 16, while
keeping the OpenMP thread count constant at 6,
and the number of MPI processes per node constant at 4,
as with previous experiments.
We excluded instances PayDomain, wikipedia, webbase, uk2005,
friendster, and twitter for these experiments as the number of non-zeros
generated by the SpGEMM did not fit into the RAM of a single compute node.
The batch size in this experiment is fixed to 8192.

Figure~\ref{subfig:spgemm-scale}
shows the relative runtime for different numbers of MPI processes.
The time that our algorithm takes per non-zero decreases
with the number of MPI processes, indicating that we do not
yet hit a performance bottleneck on our cluster.
The findings in Figure~\ref{subfig:spgemm-scale} are
supported by Figure~\ref{subfig:spgemm-breakdown}.
Local multiplication, reduce/scatter and initial send/receive rounds
scale well with the number of compute nodes.
However, broadcasting matrices takes a larger fraction
of the overall running time for higher numbers of compute nodes
(as expected).


\section{Conclusions}
\label{sec:conclusion}
%
In this paper we proposed a data structure for dynamic sparse graphs/matrices distributed over MPI processes.
This data structure allows for fast updates and redistribution. With this data structure and an adapted
communication mechanism, we designed a dynamic SpGEMM algorithm that usually performs several times faster
in practice than the static state of the art.
%
Future work could investigate replacing the SUMMA algorithm in the
3D SpGEMM by Azad \etal~\cite{doi:10.1137/15M104253X} by our algorithm to obtain a new dynamic one
with further improved communication volume.

\ifthenelse{\boolean{blinded}}
{}
{\paragraph*{Acknowledgments}
This work is partially supported by
German Research Foundation (DFG) grant GR 5745/1-1 (DyANE)
and DFG grant ME 3619/4-1 (ALMACOM).
}

\bibliographystyle{IEEEtran}
\bibliography{references,references-dyane}


\end{document}